# A CT GEOMETRY WITH MULTIPLE CENTERS OF ROTATION FOR SOLVING SPARSE VIEW PROBLEM

Jiayu Duan, Yang Li, Jianmei Cai, Xuanqin Mou[*1]

Institute of Intelligent Computing and Data Communication, Xi'an Jiaotong University, Xi'an, China

**Abstract** With the emergence of CNT (Carbon nanotube), static and instant CT scanning becomes possible. By transforming the traditionally rotated thermal source into a static ring array source composed of multiple CNTs, the imaging system can achieve high temporal resolution in scanning. However, due to the non-negligible packaging size of CNTs, the static CT based on CNTs faces sparse view problem, which affects the image quality by introducing streak artifacts. In this study, we based on the local correlation equation (LCE) to address the sparse view problem of static CT. The LCE is a series of partial differential equations (PDEs) to describe the local correlation of Radon transform in a neighborhood projection domain. Based on LCE, we analyze the characteristic of sparse view projection and propose a scanning geometry with multiple rotation centers, which is different from existing CT devices that acquires the projection around one rotation center. Specifically, in the proposed scanning geometry, the circular ring array X-ray sources is divided into several arcs while the sources of each arc share one rotation center. All rotation centers of the arcs are uniformly distributed on a small circle. The optimal distribution of the rotation centers can be optimized by the radius of the circle. Moreover, to elevate the image quality under the sparse view reconstruction, we employed the LCE to interpolate unmeasured projections. Compared to the single rotation center scheme used in existing CT geometries, the multiple rotation centers scan contributes to a more even projection distribution with same view number. The simulated results demonstrated the efficiency and potential applications of the proposed method in static CT reconstructions.

## 1 Introduction

With the characteristic of non-destructive inspection, computed tomography (CT) has been widely used in industrial inspection and medical diagnosis, etc. Under the guidance of central slice theorem, the design of CT scan geometry follows single rotation center. And to maintain the image quality of reconstruction, the number of acquisitions should be sufficient. Although with advanced fabrications, CT imaging system with thermal X-ray source now can achieve relatively high temporal resolution, e.g., 3-4 circles/second for diagnosis CTs, the bulky and heavy thermal X-ray source still holds back the further development of high temporal and spatial resolution CT.

Recently, carbon nanotube cold cathode has aroused much attention for its small size and instant response[1], which can be used to design high temporal and static scanning trajectory by replacing thermal X-ray source with CNT ring. However, due to non-neglectable interval between adjacent CNTs, the scanning faces sparse view problem, which degrades quality of the reconstructed images.

To alleviate the artifacts, some strategies can be used by taking interpolation of the sparse data[2, 3]. Introducing compressed sensing and deep networks into the reconstruction algorithms have also brought significant progresses for solving the sparse view problem[4-8]. In this way, image sparsity priors are employed in the iterative reconstruction algorithms to alleviate the artifacts, such as total variation (TV)[4], dictionary learning[5], low rank[6]. Benefited from deep learning (DL) technique, researchers directly performed CNNs in degraded reconstructions to recover the details[7]. Besides, researchers also embedded CNNs in iterative process to replace the traditional regularization, such as work in [8]. Moreover, DL techniques also help to interpolate on the sparse-view projections efficiently, which breeds the methods in projection domain and hybrid domain to iteratively solve the sparse-view problem[8].

Generally, the key issue of solving interpolation on signal is to correctly model the signal. Hence, it is a good solution to explore local data consistency condition (DCC) of Radon transform which can be considered as a signal model in projection domain. The local DCC is usually expressed as a partial differential equation (PDE). One local DCC model is John equation that has been widely used in cone beam data interpolation[9, 10]. But it needs to follow certain trajectory constraints and hence is hard to use generally.

Another local DCC is to explored the inner relationship between detector and angle based on the Radon transform [11]. In [12], this DCC was used to directly reconstruct CT gradient image. However, for dependence on the imaging object, it is hard to solve sparse view interpolation problem.

In previous work, we proposed an image independent DCC, namely local correlation equation (LCE), and transformed sparse view interpolation problem into a PDE solution [13]. We used the finite difference method to discretize the LCE equation and constructed a linear object function to solve the LCE, with the known sparse view projections serving as boundary conditions. In the case of finite differences, the more uniform the distribution of the boundary conditions, the higher the accuracy of the PDE solution. However, in existing CT scanning geometry with a single rotation center, the sparse view sampling faces the problem of uneven sampling (See Fig. 2 in II.B).

In this study, we propose a CT geometry with multiple centers of rotation to alleviate the sparse view problem by offering a better boundary condition for PDE solution. Specifically, in the proposed scanning geometry, the circular ring array X-ray sources is divided into several arcs while the sources of each arc share one rotation center. All rotation centers of the arcs are uniformly distributed on a small circle. The optimal distribution of the rotation centers can be optimized by the radius of the circle. After optimization, the distribution of projections is more even compared with traditional geometry of single rotation center. Moreover, to further elevate the image quality of the sparse view reconstruction, we employed the LCE to interpolate unmeasured projections. The results in this study have validated the efficiency of the multiple rotation centers strategy, which shows that the strategy can be a promising option for designing static CT reconstruction.

---
[1] This work was supported by the National Key Research and Development Program of China (Nos. 2022YFA1204203, 2016YFA0202003) and National natural Science Foundation of China (NSFC) (No. 62071375).
Correspondence: xqmou@mail.xjtu.edu.cn (X. Mou)

The remainder of this paper is organized as follows. Section II describes the method. In Section III, we detail the experiment arrangement and the evaluation method. In Section IV, experimental results are presented. We discuss some related issues of this study and conclude this article in Section V.

## 2 Methods

### A. Preliminary: LCE

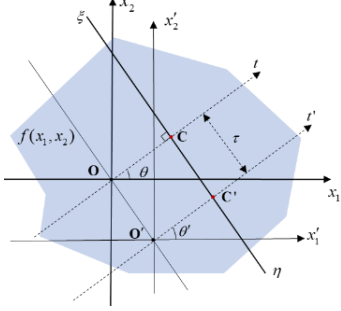

Fig. 1. The pseudo rotation centers illustration

This section we brief the image independent LCE shown in [13]. Recall the local DCC described in [11] which can be denoted as:

$$\frac{\partial}{\partial \theta_k} Rf(\boldsymbol{\theta},t) = -\frac{\partial}{\partial t}(R(x_k f))(\boldsymbol{\theta},t), k=1,...,n, \quad (1)$$

where $Rf$ denotes the Radon transform over the hyperplane perpendicular to $\boldsymbol{\theta}$ with signed distance $t$ from the origin. In 2D parallel-beam geometry, Eq.(1) can be rewritten as[12]:

$$\frac{\partial}{\partial \theta} Rf(\theta,t) = \frac{\partial}{\partial \theta_1} Rf(\theta,t)\frac{\partial \theta_1}{\partial \theta} + \frac{\partial}{\partial \theta_2} Rf(\theta,t)\frac{\partial \theta_2}{\partial \theta}$$
$$= \frac{\partial}{\partial t}(R(x\cdot\boldsymbol{\theta}^\perp)f))(\boldsymbol{\theta},t) \quad ,(2)$$

where $\boldsymbol{\theta}^\perp = (\sin\theta, -\cos\theta)$ is the unit normal vector to $\boldsymbol{\theta}$, $\boldsymbol{\theta}^\perp \cdot \boldsymbol{\theta} = 0$. From Eq.(2), we can observe that there is an implicit relationship between angle and detector differentials, which is object depended and hence hard to solve. To obtain direct relationship between differentials of the Radon variables, in [13], we introduced two pseudo-rotation centers to transform the implicit relationship between the angle and detector to an explicit relationship, which shown in Fig. 1. And the LCE can be derived as:

$$\sum_{k=0}^{n} \frac{n!}{k!(n-k)!} \frac{\partial^n Rf(\theta,t)}{\tau^k \partial t^{n-k}\partial \theta^k} = \frac{1}{\tau^n} \frac{\partial^n R'f(\theta',t')}{\partial \theta'^n} \quad (3)$$

where $Rf(\theta,t)$ and $R'f(\theta',t')$ represent the Radon transform of the same line integral under C and C' rotation center with $\theta = \theta'$ and $t = t'$, respectively.

Specifically, in circular fan beam geometry, the 1st order of LCE abbreviated as cLCE can be described as [13]:

$$\frac{\partial R}{\partial t} = \frac{1}{\tau}\left(\frac{\partial R}{\partial \omega} - \frac{\partial R}{\partial \theta}\right) \quad (4)$$

where $\omega$ is the coordinate of the equiangular detector to replace $\theta'$ and $R$ is the abbreviated form of $Rf$.

With discretization of cLCE, sparse view interpolation problem in CT can be generalized as [13]:

$$\min_{\mathbf{R}}\left\{\|\mathbf{PR}\|_2^2 + \alpha\|\mathbf{DR} - \mathbf{R}_S\|_2^2\right\}, \quad (5)$$

where $\mathbf{P} \subseteq R^{MN \times MN}$ is the discretization form of the cLCE shown in Eq.(4). M and N denote the number of views and detector, respectively. $\mathbf{D} \subseteq R^{MN \times MN}$ is a 0-1 matrix to locate the known sparse view projection. $\mathbf{R}_S \subseteq R^{MN \times 1}$ denotes the sparse view projection that has been measured in the scanning, which can be considered as the boundary condition solving the cLCE.

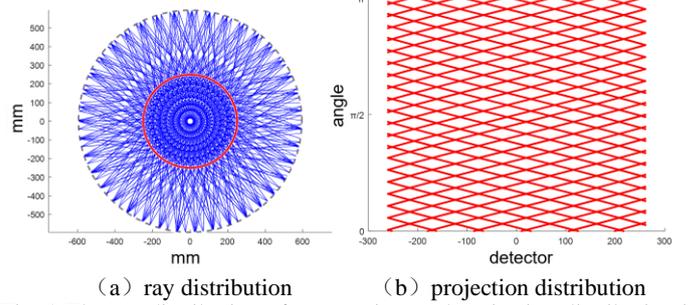

（a）ray distribution （b）projection distribution
Fig. 2 The ray distribution of sparse view and projection distribution in single rotation center geometry (1/30 sparsity), red circle denotes the FOV area

As mentioned before, solving Eq.(5) depends on the boundary condition, which depends on the distribution of known sparse view projection. In single rotation center scheme, we find that the rays' distribution of sparse view scanning is uneven and the projection exhibits dense in angle but sparse in detector, as shown in Fig. 2.

Based on above analysis, the boundary conditions of solving sparse view CT with existing geometry of single rotation center is not uniform because of the black area. The uniformity may affect the accuracy of solving cLCE. Meanwhile, we also find that the rays' distribution is dense in some areas, which is useless in elevating the performance of solving cLCE. Hence, to formularize a better boundary condition for cLCE solution, in the next, we will detail a novel scanning trajectory with multiple rotation centers.

### B. The CT geometry with multiple centers of rotation

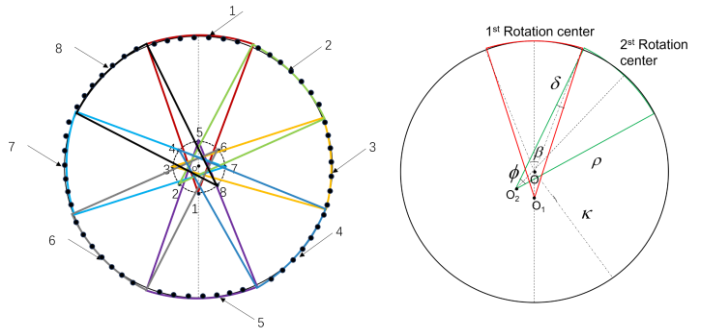

Fig. 3 The configuration of multiple centers of rotation

In the new CT geometry, we divide the whole circular ring X-ray sources into several arcs, and the number of the arcs is $S$. Each arc has the same number of X-ray sources and the sources in one arc share a fixed rotation center with the radius of $\rho$. It was worth noting that the so called whole circular ring where the array X-ray sources are arranged is not a strictly circular ring but consists of $S$ arcs that has the curvity with a little difference. The center of each arc to the origin $O$ is $\kappa$. Hence, there are $S$ rotation centers in the proposed geometry and the centers are distributed on a small circle which radius is $\rho - \kappa$.

Let us take Fig. 3 as an example to illustrate the geometry setting. The geometry contains 8 rotation centers, indexed as 1-8, and the rotation centers are evenly distributed on the circle. $O, O_k$, $k=1,2,...,8$ are the original rotation center and rotation center of k-th arc, respectively. The length of each arc is $\Pi$ with radius $\rho$. The central angle of each arc is $\phi$ and the angle difference between two adjacent arcs is $\delta$. The central angle of each arc regarding the origin $O$ is $\beta$. The complete multi-rotation center scanning geometry needs to meet the following conditions:

$$\beta \cdot \kappa = \phi \cdot \rho = \Pi$$
$$\beta = \phi - \delta. \quad (6)$$

In practice, $\Pi$ is usually fixed, then the only variable that needs to be optimized is $\delta$.

The optimization of rotation centers is not unique. In this study, we use binary search to optimize $\delta$. We propose two indexes for optimizing the centers. First, the angular distribution of projection should be denser and more uniform. Second, the overlapped FOV of all rotation centers should be as large as possible. To quantitively measure the uniformity of the projections, we used coefficient of variance (CV) as indicator, which can be described as:

$$CV = \frac{\delta_{sv}}{\mu}, \quad (7)$$

where $\mu$ and $\delta_{sv}$ represent mean and variance, respectively. Smaller $CV$ means more uniformity of the projections.

**C. The LCE based interpolation method with multiple centers of rotation**

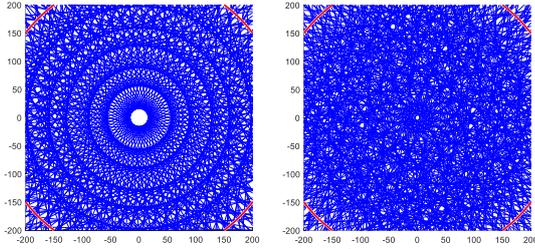

（a）single rotation center　（b）16 centers of rotation

Fig. 4 Schematic diagram of projections in the FOV. The cases of signle and multiple rotation centers are shown. The sparsity is 1/15.

Table. 1 The workflow of LCE based interpolation

| Initialization： |
|---|
| The number of rotation certer S, sparse interval M, detector size D |
| Total rays：K=M*S*D |
| for z=1:Z   Z is iteration number |
|   for k = 1:K |
|     Load k-th ray |
|     Transform into parallel index $(\theta_{para}, t)$ |
|     Calculate intersection points |
|     Find $\Delta\theta < 2\Delta\omega$, calculate (4), average $\frac{\partial R}{\partial t}$ |
|     Interpolate index $(\theta_{para}, t-1), (\theta_{para}, t+1)$ |
|   end |
|   K = 3*K |
| end |

After optimization, we set $\delta = 0.0112$ rad and the rays' distribution is shown in Fig. 4. The projection in the geometry of multiple centers of rotation distributes more even and the intersection angle between two rays are smaller. These merits benefit the solution based on cLCE.

Different from single rotation center, the situation of rays' intersection is much more complicate and may fail to use the unified interpolation in [13]. In this study, to occupy more discretization form of LCE, we directly select intersected rays with small intersection angles. If the angle between the intersected rays is less than $2\Delta\omega$, the ray is used to calculate interpolation. We take multiple interpolations, the calculated rays generated previously will be used as known rays in each iteratiion to continue to expand the interpolation. The whole workflow is listed in Table. 1.

## 3 MATERIALS

To validate the proposed method, a simulated Forbild phantom and real abdomen data from Mayo clinic are chosen [14]. The reconstruction size is set as 512×512. In single rotation center, we set the distance between source and rotation center as 570mm. The total projection number is 1024 with 672 detectors. And the detector size is set as 1mm×1mm. In multiple centers of rotation, we set S=16 and others are consistent with single rotation center.

To test the validity of the proposed method, we produced sparse view projections at two levels of sparsity. Specifically, we generated sparse views of 1/15 and 1/10 respectively for the Forbild phantom and the real abdomen data. After the interpolation is completed, the reconstructed image is obtained using ART. To reduce the discretization error of cLCE, we add TV regularization to constrain the reconstructed image. At the same time, we also compare the difference between the reconstructed images under traditional and proposed geometries without TV constraint. To quantitively evaluate the performance. In comparisons, we use PSNR as criteria.

## 4 RESULTS

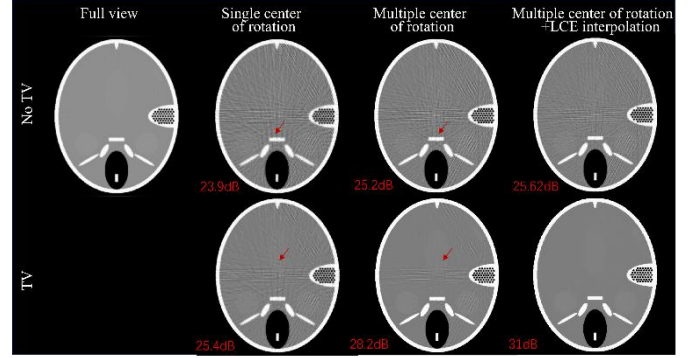

Fig. 5 Forbild phantom result with 1/15 sparsity. The first row is the iterative reconstruction result without TV regularization, and the second row is that with TV regularization. From left to right, they are the full-view reconstructed image, the single center of rotation reconstructed image, and multiple centers of rotation reconstructed image and multiple centers of rotation with LCE interpolation results, the display window level is [0.005 0.025]

Table. 2 The regularization parameter used in each case

|   | Single center of rotation | multiple centers of rotation | multiple centers of rotation+LCE |
|---|---|---|---|
| β | 0.035 | 0.065 | 0.125 |

Fig. 4 depicts the optimization result. Despite of the visualization difference, in the single rotation center geometry, the CV coefficient is 0.5124, and in the optimized multi-rotation center scanning geometry, the CV coefficient is 0.3342, indicating that the projection data is more evenly distributed in angles and detectors based on the multiple centers of rotation geometry.

To ensure the accuracy and efficiency of the algorithm, we set iteration number Z=2. In simulated phantom, two reconstruction methods were selected. The first is the reconstruction method without TV regularization, which is used to compare the worst results of reconstructed images under the proposed and traditional scanning settings. The second is to use employ TV to improve image quality. The regularization parameter of each case is listed in Table. 2. However, to ensure the resolution of the cochlear part of the Forbild phantom, the selection of TV focuses on smaller values, which may result in the strip artifacts not being fully removed.

Fig. 5 shows the results of Forbild phantom under 1/15 sparsity. The first row is the iterative reconstruction result without TV regularization, and the second row is that with TV regularization. From left to right, there are the full-view reconstructed image, reconstructed image with single center of rotation, the reconstructed image with multiple centers of rotation and the reconstructed image with multi-rotation center and cLCE interpolation, respectively. As we can see, the proposed multiple centers of rotation geometry can effectively reduce the streak artifacts caused by sparse view even without interpolation, as indicated by the red arrows. With the cLCE interpolation, the image quality is better improved. Specifically, by using TV constraint, the result is ideal in visual.

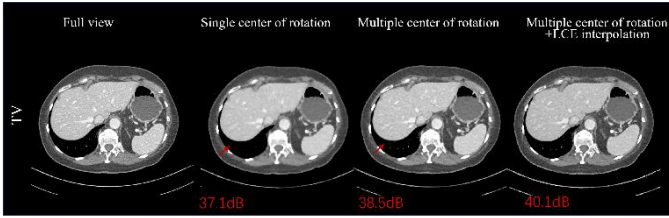

Fig. 6 Abdomen result with 1/10 sparsity. The figure arrangement is the same as Fig. 5. The display window level is [0.008 0.025]

While the visual results are satisfied, the PSNR index improvements regarding the cases without TV constraint are not obvious. We owe to that in applying cLCE on the multiple centers of rotation geometry, there has not been an ideal differential computation scheme to shun discretization errors. However, TV regularization can effectively suppress the errors and achieves obvious PSNR improvement.

In real abdominal data, we only discuss the results under TV regularization for aforementioned reasons. Fig. 6 shows the reconstruction results of abdominal data with 1/10 sparse scanning. From the results, it is evident that under the geometry of a single rotation center, the uneven sampling leads to the loss of small details in the region of lung. However, under the multiple centers of rotation geometry, fine details can be effectively preserved; furthermore, with the cLCE interpolation, the quality of the images can be further improved.

## 5 DISCUSSION AND CONCLUSION

In this study, we follow our previous image independent PDE, namely LCE to solve the sparse view problem in static CT. Specifically, inspired by the LCE, we propose the scanning trajectory of multiple rotation centers. With the novel geometry, the reconstruction quality can be compensated by the adjacent projections during the iterations. The results show the efficiency of the proposed trajectory. This new CT scanning can be easily applied to cold cathode X-ray array based geometry to solve the sparse view problem.

Fig.7 shows two sketches of the multi-center imaging system based on the arc and line distributed CNTs. Moreover, with multiple centers of rotation, each source can just cover a part of object, which reduces the radiation dose and volume size of imaging system. In our previous work, a flat panel source was successfully employed to realize a CABI tomosynthesis scheme that is also a multi-center imaging system [15].

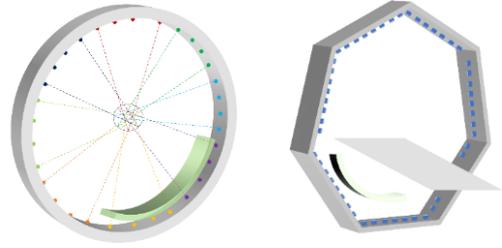

Fig. 7 The examples of multiple centers of rotation geometry

Although the results are promising, it still faces some problems. First, the choice of the multiple rotation centers should be further studied. How to design a fine objective function or deduce an analytic solution for rotation centers selection will be our future work. Second, the current LCE interpolation method is primary, which may degrade the performance of applying LCE. Hence, in the future, we will introduce deep learning or other technics to realize the LCE based interpolations.

## References

[1] C. Inscoe, Y. Z. Lee, J. Lu, and O. Zhou, "Development of CNT X-ray technology for medical and dental imaging," in Nanostructured Carbon Electron Emitters and Their Applications: Jenny Stanford Publishing, 2022, pp. 269-288.
[2] H. Lee, J. Lee, H. Kim, B. Cho, and S. Cho, "Deep-neural-network-based sinogram synthesis for sparse-view CT image reconstruction," IEEE Transactions on Radiation and Plasma Medical Sciences, vol. 3, no. 2, pp. 109-119, 2018.
[3] M. Kalke and S. Siltanen, "Sinogram interpolation method for sparse-angle tomography," Applied Mathematics, vol. 2014, 2014.
[4] E. Y. Sidky, C.-M. Kao, and X. Pan, "Accurate image reconstruction from few-views and limited-angle data in divergent-beam CT," Journal of X-ray Science and Technology, vol. 14, no. 2, pp. 119-139, 2006.
[5] Q. Xu, H. Yu, X. Mou, L. Zhang, J. Hsieh, and G. Wang, "Low-dose X-ray CT reconstruction via dictionary learning," IEEE transactions on medical imaging, vol. 31, no. 9, pp. 1682-1697, 2012.
[6] K. Kyungsang et al., "Sparse-View Spectral CT Reconstruction Using Spectral Patch-Based Low-Rank Penalty," IEEE Transactions on Medical Imaging, 2015.
[7] C. Hu et al., "Low-Dose CT With a Residual Encoder-Decoder Convolutional Neural Network," IEEE Transactions on Medical Imaging, 2017.
[8] G. Wang, J. C. Ye, and B. D. Man, "Deep learning for tomographic image reconstruction," Nature Machine Intelligence, vol. 2, no. 12, pp. 737-748, 2020.
[9] F. John, "The ultrahyperbolic differential equation with four independent variables," Fritz John: Collected Papers Volume 1, pp. 79-101, 1985.
[10] S. K. Patch, "Computation of unmeasured third-generation VCT views from measured views," IEEE transactions on medical imaging, vol. 21, no. 7, pp. 801-813, 2002.
[11] F. Natterer, The mathematics of computerized tomography. SIAM, 2001.
[12] S. Tang, X. Mou, J. Wu, H. Yan, and H. Yu, "CT gradient image reconstruction directly from projections," Journal of X-ray Science and Technology, vol. 19, no. 2, pp. 173-198, 2011.
[13] X. Mou and J. Duan, "Exploring the redundancy of Radon transform using a set of partial derivative equations: Could we precisely reconstruct the image from a sparse-view projection without any image prior?", arXiv:2405.19200v2, Aug 2024. [Online]. Available: https: //doi.org/10.48550/arXiv.2405.19200
[14] C McCollough. Tu-fg-207a-04: Overview of the low dose ct grand challenge. Medical physics, 43(6Part35):3759–3760, 2016.
[15] J. Duan, Y. Li, S. Kang, et al., "Coded Array Beam X-ray Imaging Based on Spatio-sparsely Distributed Array Sources." Fully3D2023, 2023, pp. 86-89."